# Giant magneto-impedance in Ag-doped $La_{0.7}Sr_{0.3}MnO_3$

S.K.Ghatak and B.Kaviraj[a]
Department of Physics & Meteorology, Indian Institute of Technology, Kharagpur 721302, India

T. K. Dey
Cryogenic Engineering Center, Indian Institute of Technology, Kharagpur 721302, India

## Abstract

The resistive and reactive parts of the magneto-impedance of sintered ferromagnetic samples of $La_{0.7}Sr_{0.3-x}Ag_xMnO_3$ (x = 0.05, 0.25) have been measured at room temperature ($<T_c$) over a frequency interval from 1KHz to 15MHz and in presence of magnetic field up to 4KOe. The field dependence of relative change in resistance ($\Delta R/R(0)$) is small in KHz region but increases strongly for higher frequency of excitation. The maximum value of ($\Delta R/R(0)$) at H =4KOe and for $\omega$ =15MHz is around 70%.On the contrary the corresponding change in reactance ($\Delta X/X(0)$) has less frequency sensitivity and the maximum occurs at $\omega \approx$1MHz. The magneto-impedance is negative for all frequencies. The normalized magneto-impedance $\delta Z$ as defined by [Z(H)-Z(0)]/[Z(0)-Z(4K)] when plotted against scaled field $H/H_{1/2}$ is found to be frequency independent ; $H_{1/2}$ is the field where $\delta Z$ is reduced to half its maximum. A phenomenological formula for magneto-impedance, Z (H), in a ferromagnetic material, is proposed based on Pade' approximant. The formula for Z (H) predicts the scaled behavior of $\delta Z$.

[a] **Corresponding author**

**Email address: bhaskar@phy.iitkgp.ernet.in**

## I. Introduction

Mixed manganites $La_{1-x}A_xMnO_3$ (A= Ca, Sr, Ba) are compounds of great interest due to their colossal d.c magneto-resistance (CMR) behavior. These compounds undergoes a transition from paramagnetic–semiconducting to ferromagnetic–metallic phase at critical temperature $T_c$ [1,3]. Some of these systems exhibit a mixed phase ground state, characterized by magnetic and electronic phase separation (PS)[2]. A complex interplay between spin, charge and lattice is the characteristic feature of manganites and is considered to be the paramount reason for CMR effect. The CMR effect is normally observed near to $T_c$ and in presence of moderately large magnetic field. Doped manganites are inhomogeneous in nature[2,4] and the degree of inhomogeneity depends on number of factors like preparation condition, structural inhomogeneity (polycrystalline), doping level etc. The effect of homogeneity on magnetic and transport properties of manganites is usually minimal at optimal doping. Needless to say, all these factors lead to the magnetic inhomogeneity that in turn influences transport. As the magnetic field can reduce magnetic inhomogeneity, the field dependence of resistance in single crystalline phase differs from that of the polycrystalline state. In particular a large drop of resistance is observed at low magnetic field and in the ferromagnetic state. This low field magneto-resistance (MR) is associated with spin dependent charge transport across the grain boundaries[5]. The disorder in the magnetic state across boundary interface can be removed by small magnetic field and this in turn reduces spin-disorder scattering. The magnetic inhomogeneity at grain boundary can hinder the flow of a.c current and a drop of impedance is then expected due to partial suppression of hindrance by the magnetic field.

The magneto-impedance (MI) in these systems is less explored in comparison to the d.c counterpart.

The giant MI in powdered $La_{0.7}Ba_{0.3}MnO_3$ has been first observed in microwave region with a large drop in absorption in presence of small field ~ 600 Oe[6]. Subsequently, the MI in this frequency domain has also been reported in similar manganite compounds[7]. It was noted that the magnetic inhomogeneity plays a crucial role for the giant MI effect in microwave frequency[7]. The magneto-impedance in $La_{0.67}Ba_{0.33}MnO_3$ in MHz region has been reported and relative change in impedance for magnetic field < 100 Oe is found to be small[8]. The giant magneto-impedance (GMI) phenomenon is normally investigated in soft ferromagnetic metallic systems that include crystalline and amorphous state. The ferromagnetic materials like the transition metal-metalloid amorphous alloys are forerunners in exhibiting large negative MI at low frequency[9-12]. The Fe-or Co-based metallic glasses in the form of ribbon or wire are magnetically soft materials and the relative decrease in impedance in presence of small field $\geq 50$ Oe is very large[9-13]. The qualitative understanding of MI effect is based on the magnetic response of ferromagnetic material to the a.c. field in presence of a biasing d.c. field, and the dynamic permeability $\mu(\omega,H)$ is the measure of the response. The impedance of a metal depends on skin depth which in turn depends on $\mu$ in ferromagnetic state. The dynamic permeability can be reduced to a large extent in soft ferromagnetic materials in presence of small magnetic field resulting in a large drop of impedance[14-15]. In this paper, we report an a.c response of polycrystalline sample of $La_{0.7}Sr_{0.3-x}Ag_xMnO_3$ (x = 0.05, 0.25) at room temperature (< $T_c$) over a frequency interval upto 15MHz. The resistive and reactive components of magneto-impedance (MI) are measured in presence of biasing d.c. magnetic field (H) upto 4KOe. Both components of MI decrease sharply with field H and tend to saturate within this field . The maximum value of MI increases with frequency of excitation and is of order of 65% at H = 4KOe for 15MHz. The data for different frequencies collapse into a single curve when reduced MI, scaled by its maximum value, is plotted in terms of scaled magnetic field. A phenomenological formula for impedance is proposed to describe the field dependence and collapse of MI data at different frequencies.

## II. Experimental

Powder samples of $La_{0.7}Sr_{0.3-x}Ag_xMnO_3$ (x = 0.25, 0.05) were prepared through thermolysis of an aqueous, polymeric-bond precursor solution starting from water-soluble coordinated complexes of the composing metal ions. $La_2O_3$, $Sr(NO_3)_2$, $AgNO_3$ in appropriate molar ratios are dissolved into dilute nitric acid to obtain the stocks of respective metal nitrates. Separately, aqueous solution of the $Mn(CH_3COO)_2$ tetra-hydrate (99% purity) (0.01M) were prepared by dissolving stoichiometric amounts of the salt in distilled water. The nitrate stocks and the manganous acetate solution were mixed and heated to about $80^0$C. Into the hot solution, calculated amount of Poly-vinyl alcohol (PVA), Tri-ethanol amine (TEA) and Sucrose were added to obtain the final stock solution. Appropriate amount of the same was then taken to heat over a hot plate (~$200^0$C) with continuous stirring and was eventually evaporated to dryness. On complete evaporation, the precursor solution gave rise to a fluffy charred organic mass that had the desired metal ions embedded in its matrix. The carbon rich mass was crushed to constitute the precursor powders. Subsequent calcinations of the precursors at $800^0$C (6hrs) resulted in the carbon-free, nano-sized powders having the desired compositions.

Pellets prepared from this precursor powders have been heat treated in air at $1050^0$C for 24 hrs and was then furnace cooled to room temperature.

XRD of the samples confirmed single phasic nature of all the prepared samples. X-ray diffraction patterns have been indexed in the rhombohedral system with space group $R\bar{3}C$ . The calculated lattice parameters are listed in the Table I. The crystallite size of the prepared samples, estimated by Debye-Scherer method lies between 30-40nm. The Curie temperature was determined from the temperature dependence of the a.c susceptibility and is $366.3^0$K and $303.6^0$K for x = 0.25 and 0.05 respectively[16].

The sample of approximate cylindrical in shape (8mm long and 2mm diameter) is used for the measurement of impedance. A signal coil of 25 turns was wound over the midpoint of the sample and the impedance of the coil with and without the sample was measured with Impedance analyzer (Agilent 4294A) at room temperature ($298^0$K) .The biasing d.c. field 'H' upto 4KOe was applied parallel to the exciting a.c. field. The resistive and reactive components of impedance were measured upto 15MHz using excitation current of 20mA amplitude resulting in an a.c. field ~ 0.9 Oe. The magnetization of the sample was measured using a.c. magnetometer and the maximum magnitude of magnetic field was 450Oe at 70Hz frequency.

## III. Results & Discussions:

### A. $La_{0.7}Sr_{0.3-x}Ag_xMnO_3$ (x = 0.25)

The resistance 'R' and reactance 'X' of the sample are displayed in Fig. 1 as functions of frequency for three different bias fields H applied parallel to the axis of the sample. The a.c. excitation field is also parallel to H.

At low frequency both components of impedance varies little with frequency and magnetic field. Above frequency around 100KHz, R and X varies strongly with frequency. The reactance is higher than resistance for higher frequency (upto 15MHz). Note that the frequency is given in logarithmic scale. The decrease in R and X with magnetic field is also large at higher frequency. The results for the field variation of relative change of resistance and reactance as defined by '$\delta R$' = [(R(H) – R(0))/ R(0)]% and '$\delta X$' = [(X(H) – X(0))/ X(0)]% are displayed in Fig. 2 (a) and (b) respectively .

Both $\delta R$ and $\delta X$ are decreased in presence of field H and therefore, the material exhibits negative magneto-impedance. The drop in $\delta R$ is sharp at low fields and saturates at higher fields. The magnitude of $\delta R$ depends upon frequency of excitation and within KHz region the maximum decrease is only within 5%. At higher frequency, the maximum value of $\delta R$ becomes 35% at 1MHz and 68% at 15 MHz. In contrast to $\delta R$, stronger dependence of $\delta X$ with magnetic field is observed at lower frequency. For a given field, the variation of magneto-inductance with frequency is small within the measured range. The results for 1MHz and 0.15MHz almost coincide and maximum value of $\delta X$ is around 65%. The maximum change in $\delta X$ is a decreasing function of frequency in MHz region. This is opposite to $\delta R$ variation (Fig. 2(a)).With the increase in frequency of excitation field, the d.c. field required to saturate both components of impedance becomes higher. The field dependence of impedance $Z = [R^2 + X^2]^{1/2}$ is plotted in terms of relative change $\delta Z = [Z(H) - Z(0)]/ Z(0)$ in Fig. 3(a) for above three frequencies.

The maximum value '$\delta Z_M$' of '$\delta Z$' increases with frequency and is equal to 66% for 15MHz and 49% for 0.15 MHz. Although there is little variation of $\delta Z$ beyond 2KOe, $\delta Z_M$ is taken at a field of 4 KOe. The field where $\delta Z$ becomes half of its maximum value is different for different frequencies and is denoted as $H_{1/2}$. In order to look for a general behavior, the data in Fig. 3(a) are examined in terms of reduced units ($\delta Z/\delta Z_M$) and $H/H_{1/2}$ and the corresponding plot is shown in Fig. 3(b). It is remarkable that all data points collapse into a single curve indicating nearly frequency independent relationship between the reduced quantities ($\delta Z/\delta Z_M$) and $H/H_{1/2}$.

**B. $La_{0.7}Sr_{0.3-x}Ag_xMnO_3$ (x = 0.05)**

In this sample, the observed frequency response of R and X at different bias fields were very similar to that of the previous sample (x = 0.25) but the magnitude of zero-field components were higher in x = 0.05 compared to x = 0.25.

The field variation of the relative change in the real and imaginary components of impedance at different frequencies is shown in Fig. 4. We note the decrease of both $\delta R$ and $\delta X$ of the sample with magnetic field, H, thus exhibiting negative magneto-impedance. The variation in $\delta R$ is higher at small fields and at higher frequencies the field required to saturate $\delta R$ increases. The maximum relative change in $\delta R$ depends upon the excitation frequency and it increases up to 71% at the highest chosen excitation frequency (15MHz). In contrast to $\delta R$, $\delta X$ exhibits sharper fall with H at low frequencies. For a given field, the variation of $\delta X$ with frequency is almost negligible.

The field dependence of impedance $Z = [R^2 + X^2]^{1/2}$ is plotted in terms of relative change $\delta Z = [Z(H) - Z(0)]/ Z(0)$ in Fig. 5(a) for the above three frequencies. The field behavior of $\delta Z$ at different frequencies is governed by that of $\delta X$ at low fields. This appears from the convergence of the $\delta Z$ curves of different frequencies at low fields similar to the field dependence of $\delta X$ at different frequencies (Fig. 4(b)). This is because at low fields the variation of $\delta X$ is larger than that of $\delta R$. The maximum reduction in $\delta Z$ (denoted by $\delta Z_m$, value at H = 4KOe) increases from 51% at 0.15MHz to 66% at 15MHz. The field where $\delta Z$ becomes half of its maximum value is different for different frequencies and is denoted as $H_{1/2}$. Like the previous case, the data in Fig. 5(a) are replotted in terms of reduced units ($\delta Z/\delta Z_M$) and $H/H_{1/2}$ (Fig. 5(b)). It shows that all data points collapse into a single curve indicating nearly independent frequency relationship between the reduced quantities ($\delta Z/\delta Z_M$) and $H/H_{1/2}$.

The magnetization measurements have been carried out for both the samples by fluxmetric method at $T=300^0K$ and 70Hz frequency. The maximum applied magnetic field was 450 Oe. The initial magnetization curves have been plotted for both the samples and are shown in Fig. 6. As expected the sample with a higher $T_c$ shows larger values of magnetization (M) and also higher values of initial permeability. The initial permeability '$\chi_i$' was found to be 115 and 75 for x = 0.25 and 0.05 respectively.

In polycrystalline manganites, a sharp drop in resistance is observed at low fields and a much slower variation in the higher field region. The negative magneto-resistance (MR) has a larger magnitude at lower temperatures. This MR in polycrystalline state is dominated by intergrain effects [10]. The grain boundary creates local disruption in magnetic order, and this magnetic disorder is important in charge transport due to high degree of spin polarization in these oxides. Thereby, the intergrain effects on charge transport appear to be similar to that associated with rotation of the magnetic domain in ferromagnetic crystal. The high field MR is related to spin alignment within the grain. Inside magnetic domain the charge carrier can easily be transferred between pairs of $Mn^{+3}$ and $Mn^{+4}$. On the other hand the carriers are scattered due to spin disorder across the grain boundary. On application of a moderately low field the spins in the domain associated with grain boundary can be aligned into collinear configuration and thereby negative MR results. This picture can also be considered for qualitative description of the results of magneto-impedance. Based on classical electrodynamics the magneto-impedance effect is usually described in terms of the screening of electro-magnetic field in magnetic metallic systems[5-8]. In presence of small exciting a.c. magnetic field the magnetic response measured by permeability ($\mu$) plays a crucial role in determining the magnitude of the field penetration depth,[9] $\delta$, as $\delta$ varies as $\mu^{-1/2}$ in magnetic metallic materials. The skin depth decreases to a large extent due to high values of permeability for soft ferromagnets and this leads to higher values of impedance. However, this magnetic response can be modulated by a biasing d.c. magnetic field. When the d.c. magnetic field is large compared to relevant anisotropy field, the permeability drops to a low value due to nearly complete alignment of all spins parallel to biasing field. Hence the skin depth becomes large and the impedance goes down resulting in a large MI effect. As biasing field increases, the material tends to be magnetically homogeneous .Therefore, in presence of high biasing field the response of the material to the small a.c field becomes very small and also decreases slowly with H. This small variation of Z with H is related to weaker dependency of $\mu_{eff}$ on H. In order to describe the reduced graph (Fig. 5(b)), a phenomenological formula is derived below using the results on field dependence of Z on H and the Pade approximant. It is observed from the high field region data of Z that dZ/dH can be approximated as $|dZ/dH| \propto H^{-2}$ .On the other hand, at very low H, dZ/dH<0. Considering these results and following the Pade approximation, we write dZ/dH as a ratio of two polynomials:

$$\left(\frac{dZ}{dH}\right) = \frac{-A - BH}{C + DH^3} \qquad \text{.................. (1)}$$

where A,B,C,D are parameters independent of H. Integrating equation (1) and with the boundary condition that magneto impedance $Z \to 0$ as $H \to \infty$, the equation for Z takes the form:

$$\frac{Z}{Z_0} - 1 = dZ = \frac{3}{2\pi}\left(\frac{\pi}{2} - tan^{-1}\left(\frac{2H - H_2}{\sqrt{3}H_2}\right)\right) + \left(\frac{H_2 + H_0}{H_2 - H_0}\right)\frac{\sqrt{3}}{4\pi} log \frac{(H + H_2)^2}{H^2 - HH_2 + H_2^{\ 2}} - 1$$

................ (2)

where $H_0 = A/B$ and $H_2 = (C/D)^{1/3}$ and $Z_0$ is zero-field value of Z.

The expression (2) relates the dependence of relative change in Z (denoted by dZ) upon three parameters, H, $H_0$ and $H_2$. Two representative plots exhibiting the dependence of dZ on field parameter H for different values of $H_0$ and $H_2$ are given in Fig. 7(a) and 7(b).

The curves in Fig. 7 show that the impedance dZ decreases monotonically with field H for low values of $H_0$ (curves with $H_0 = 0.5$ and -0.3 are depicted). The sharpness and extent of decrease in dZ depends on the values of $H_2$. For small values of $H_2$, the decrease is nearly 100% and dZ sharply falls as H increases. With the increase in $H_2$, a slower variation of dZ is found in both the cases. With higher value of $H_0$, the qualitative behavior of dZ is same. For $H_0 < 0$, instead of monotonic decrease, dZ increases as H increases from zero and passes through a maximum for small values of $H_2$ (Fig. 7(b)). However, this maximum disappears when $H_2 >> H_0$. For a given field, the amount of decrease in dZ becomes smaller for higher value of $H_2$. As the experimental data is closer to the situation with lower values of $H_0$, we examine the results of Fig. 7(a) in a different way. The results are re-plotted in terms of scaled variables $H/H_{1/2}$ in Fig. 8.

Here $H_{1/2}$ is the field where dZ becomes half of its maximum value. The results of dZ for different $H_2$ collapse into a single curve. As $H_{1/2}$ is higher for larger $H_2$ the curve for higher $H_2$ shifts to lower values of $H/H_{1/2}$ compared to the curve for lower $H_2$. The value of $H_0$ is taken to be smaller than $H_2$ in order to reproduce observed small slope of dZ near zero field.

Experimental results (Fig. 3(b) & 5(b)) for both the samples and three frequencies are superposed with the results of equation (2) in Fig 9. Different points represent the value of ($\delta Z/\delta Z_M$ ) and the solid line refers to the result of equation (2). The plot shows an excellent agreement between the phenomenological equation (2) and the experiment and it demonstrates that equation (2) can be used to describe the scaled behavior of relative Z with respect to the scaled field $H/H_{1/2}$.

## IV. Conclusions

Magneto-impedance measurements for both the samples of $La_{0.7}Sr_{0.3-x}Ag_xMnO_3$ with x = 0.25 and 0.05 suggests there is an appreciable change in the resistive part of Z with the excitation frequency and the sharpness of MI curves together with the maximum change increases at higher excitation frequencies. For a particular field, the frequency response of the change in the reactive component of Z is very minimal. An interesting result is obtained when the relative change in Z is plotted as a function of H scaled to $H/H_{1/2}$ when all the MI curves for different frequencies collapse into a single one. This shows the frequency independent nature of MI once H is scaled as $H/H_{1/2}$. A phenomenological model based on Pade approximant is presented to describe the scaled behavior of $\delta Z$ where we observe the experimental data to be very well supported by the theoretical model.

**TABLES:**

**Table I.** Structural and some physical parameters of $La_{0.7}Sr_{0.3-x}Ag_xMnO_3$ compounds

| Sample (x) | a=b (AU) | $Mn^{4+}$ (%) | Crystallite size (nm) | $T_{MI}$ (K) | $T_c$ (K) | $\rho_{300K}$ (Ω-cm) | $MR_{peak}$ (%) | $MR_{300}$ (%) |
|---|---|---|---|---|---|---|---|---|
| 0.05 | 5.5014 | 35 | 33 | 319.6 | 303.6 | 0.164 | 1.25 | 1.93 |
| 0.25 | 5.5097 | 55 | 39 | 366.3 | 363.4 | 0.036 | 16.72 | 8.76 |

## Figure Captions:

FIG. 1. Frequency variation of resistance 'R' (a) and reactance 'X' (b) of $La_{0.7}Sr_{0.3x}Ag_xMnO_3$ (x = 0.25) at H =0, 0.6, and 4 KOe which is parallel to the exciting a.c magnetic field.

FIG. 2. Variation of δR (a) and δX (b) with magnetic field H for frequency 0.15, 1 and 15MHz for the sample $La_{0.7}Sr_{0.3-x}Ag_xMnO_3$ (x=0.25).

FIG. 3. Plot of relative MI change (δZ) with magnetic field H (a) and reduced MI ($\delta Z/\delta Z_M$) with reduced field $H/H_{1/2}$ (b) at different frequencies of 0.15, 1, and 15MHz.

FIG. 4. Variation of δR (a) and δX (b) with magnetic field H at different frequencies (0.15, 1 and 15MHz) for the sample $La_{0.7}Sr_{0.3-x}Ag_xMnO_3$ (x = 0.05).

FIG. 5. Plot of relative MI change δZ with magnetic field H (a) and reduced MI ($\delta Z/\delta Z_M$) with reduced field $H/H_{1/2}$ (b) at different frequencies of 0.15, 1 and 15MHz .

FIG. 6. Magnetization as a function of magnetic field for samples $La_{0.7}Sr_{0.3-x}Ag_xMnO_3$ (x=0.25 and 0.05).

FIG. 7. Plot of dZ as a function H for $H_0$ = 0.5 (a) and $H_0$ = -0.3 (b) and at various values of $H_2$ as denoted in the plots.

FIG. 8. Re-plot of data in Fig. 7(a) in terms of reduced field $H/H_{1/2}$.

FIG. 9. Plot of $(\delta Z/\delta Z_M)$ (points) and dZ (Eqn.2) as function of $H/H_{1/2}$.

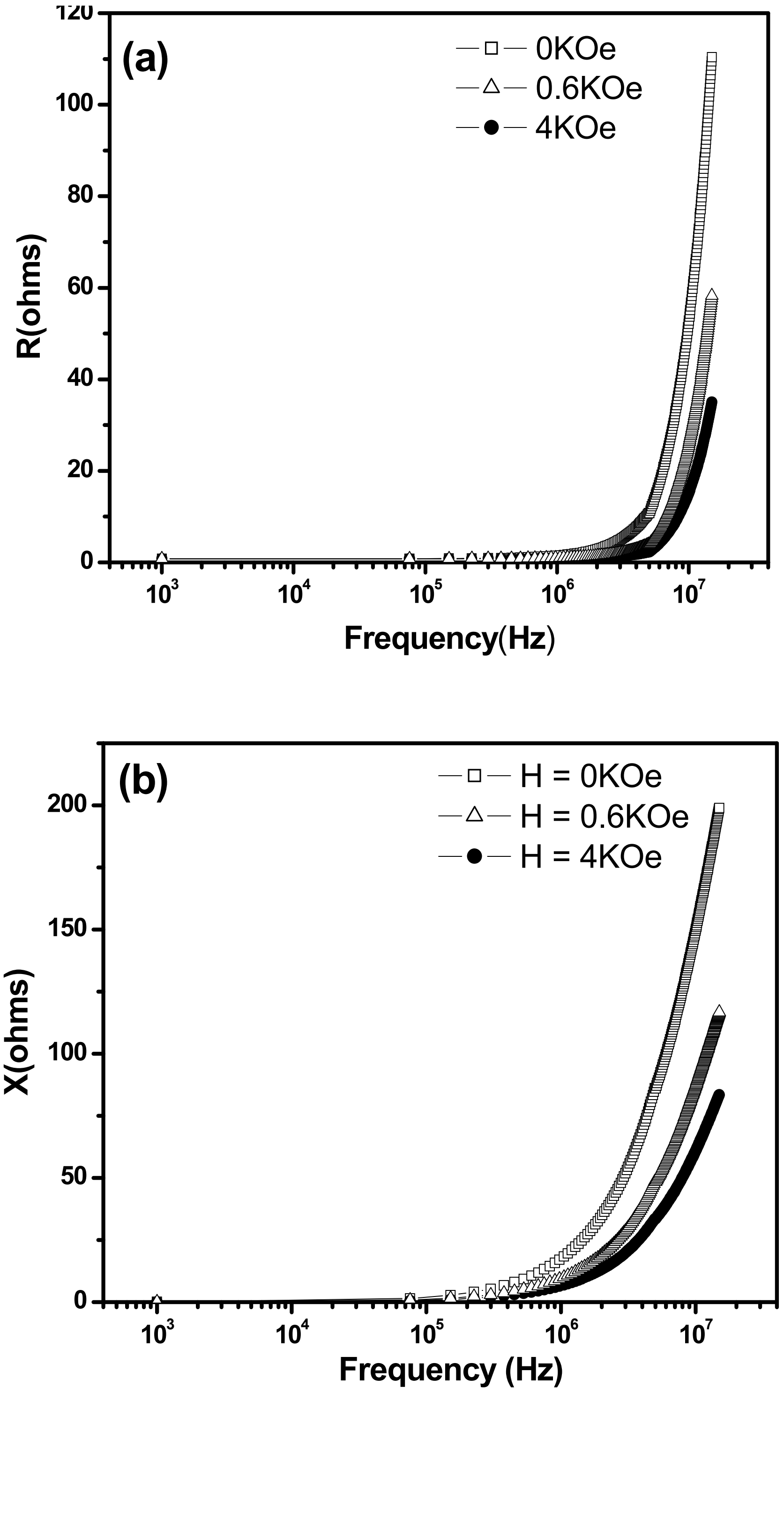

**FIG. 1 Ghatak et al.**

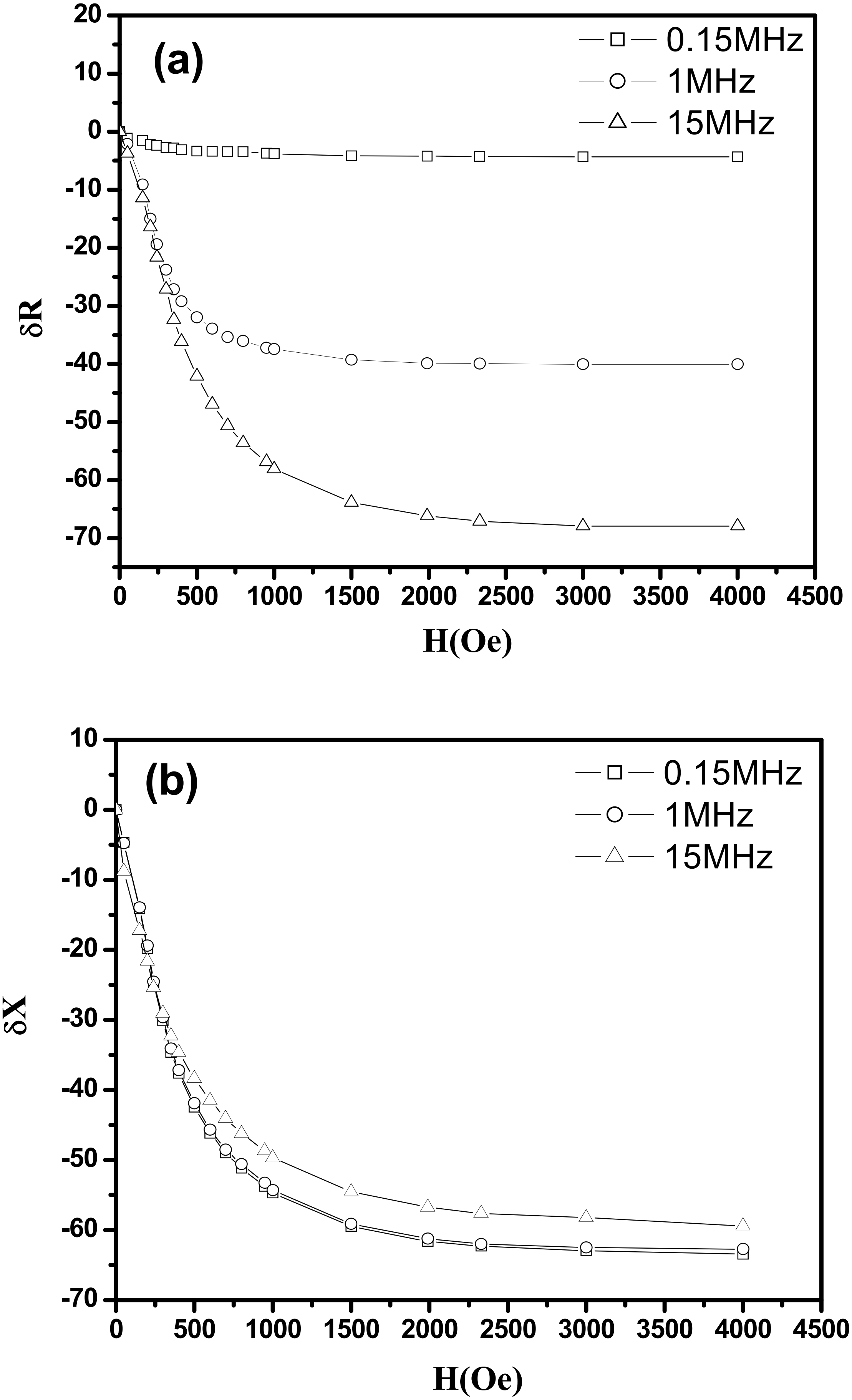

**FIG. 2 Ghatak et al.**

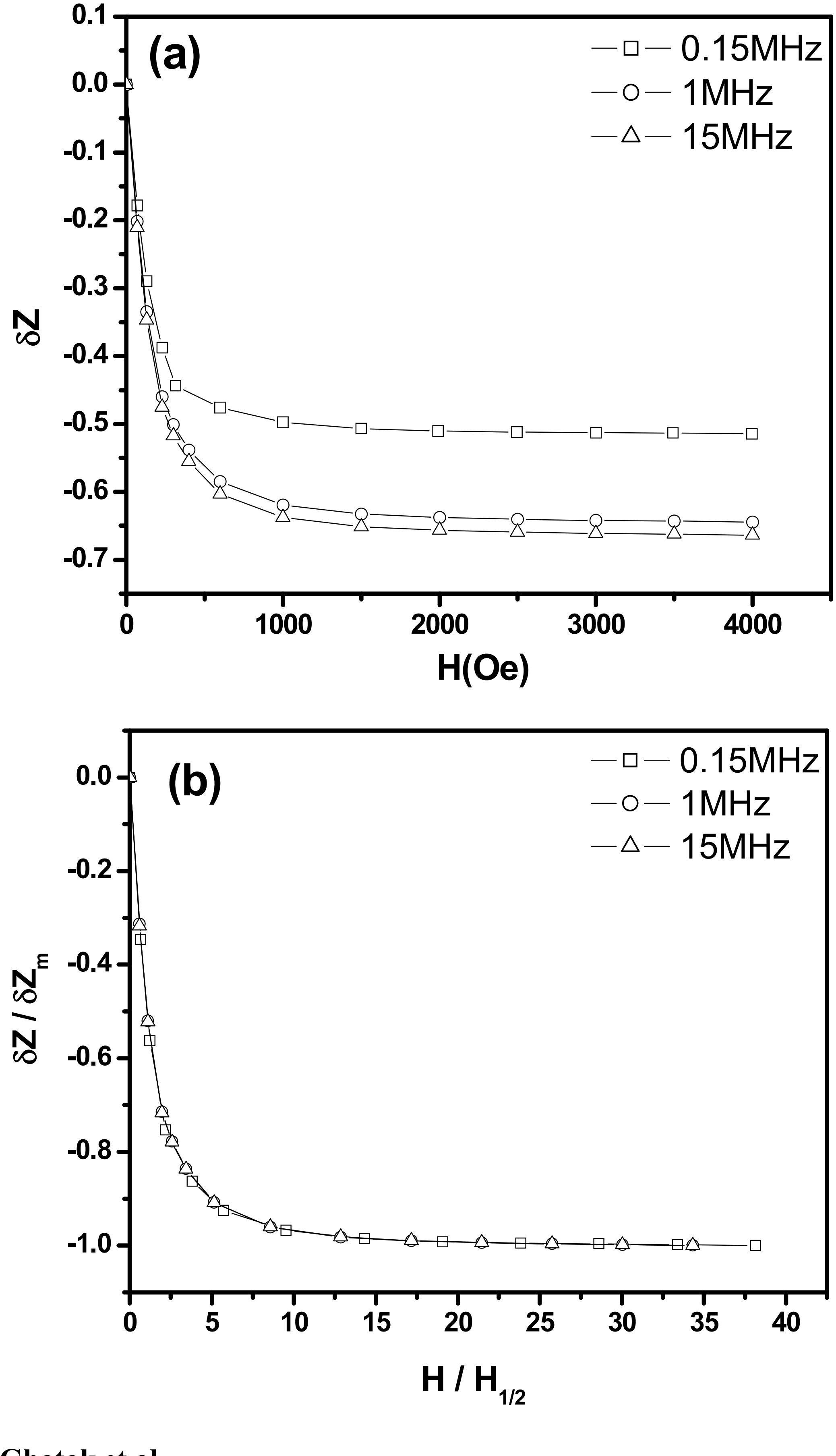

**FIG. 3 Ghatak et al.**

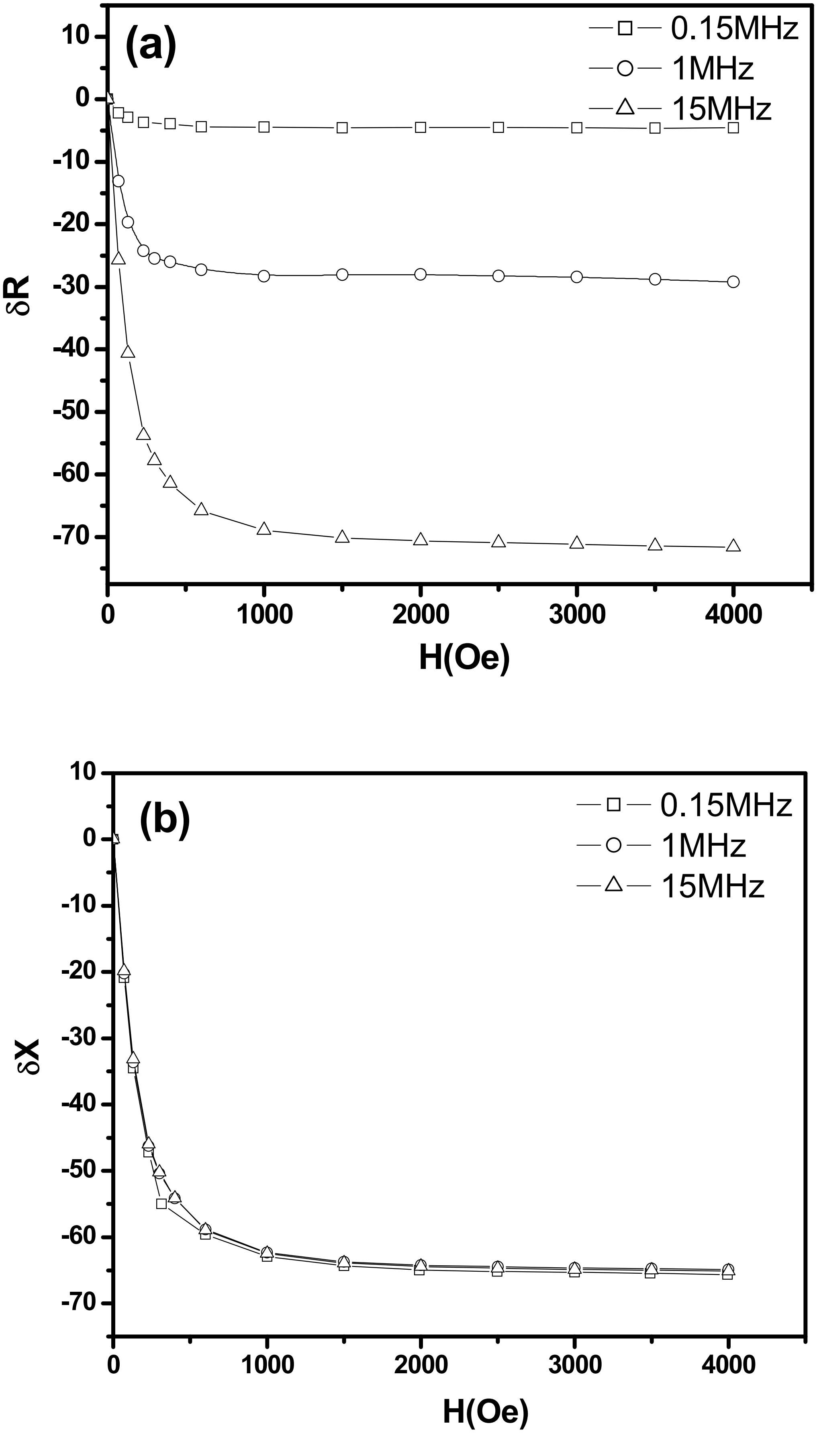

**Fig. 4 Ghatak et al.**

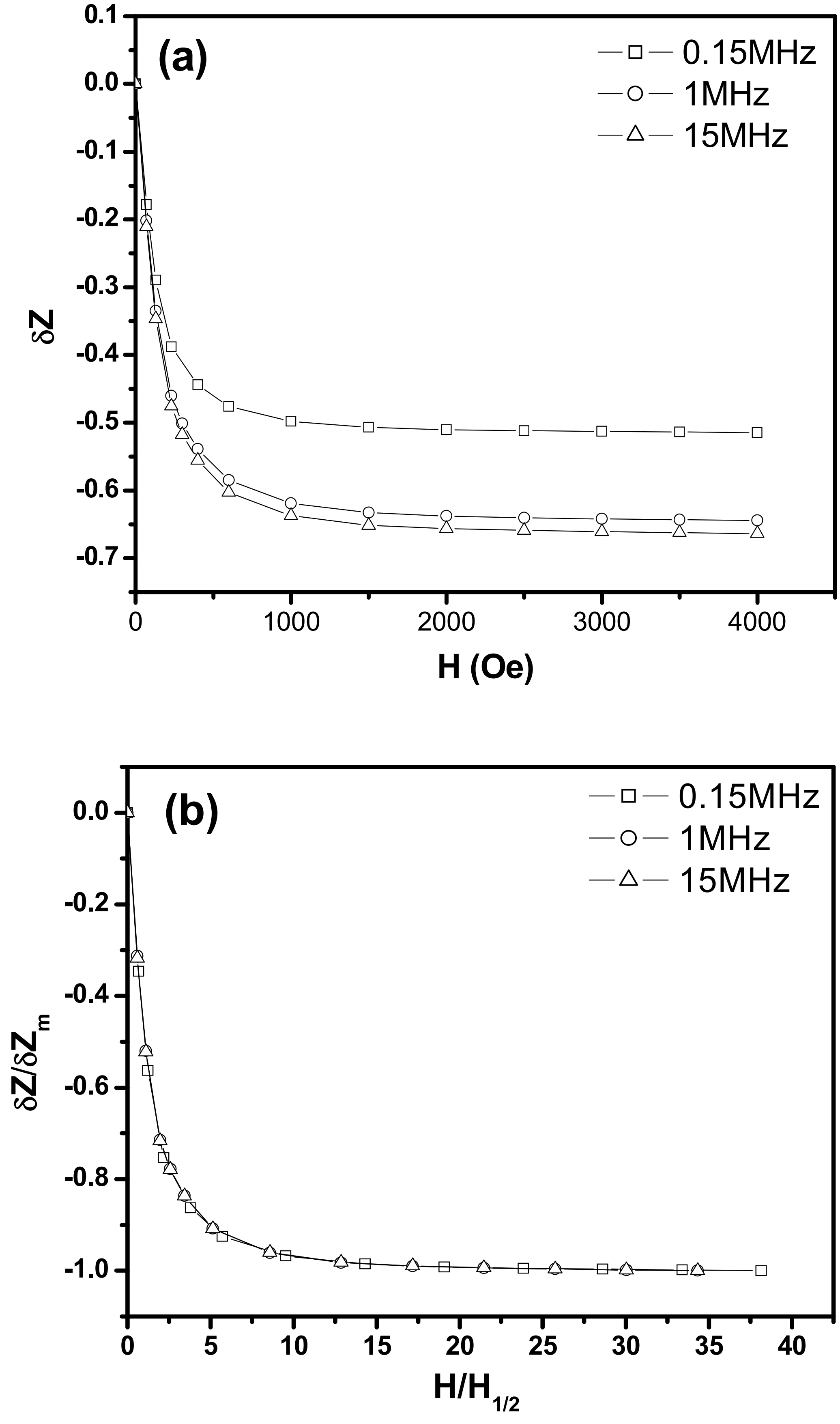

**Fig. 5 Ghatak et al.**

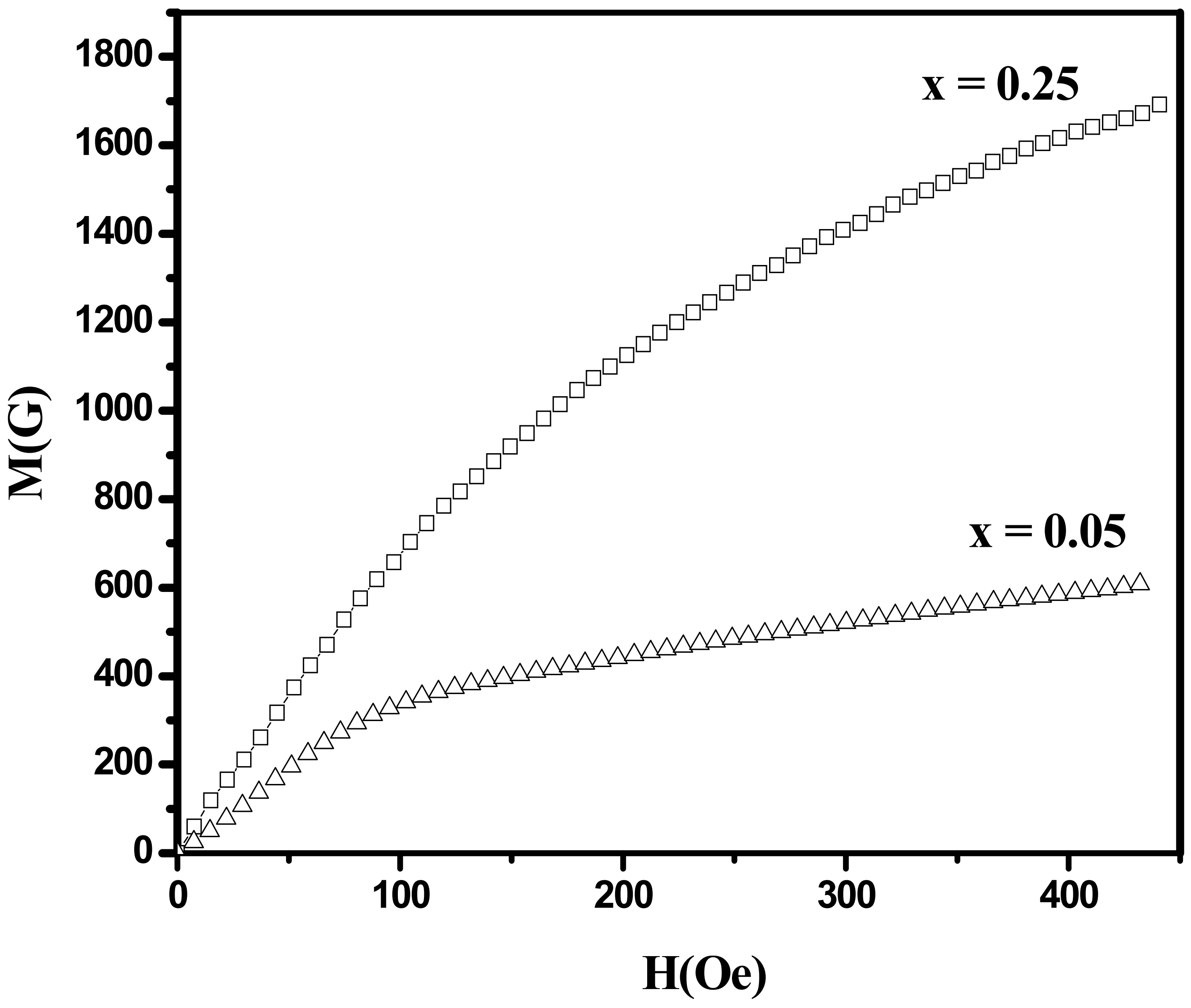

FIG. 6 Ghatak et al.

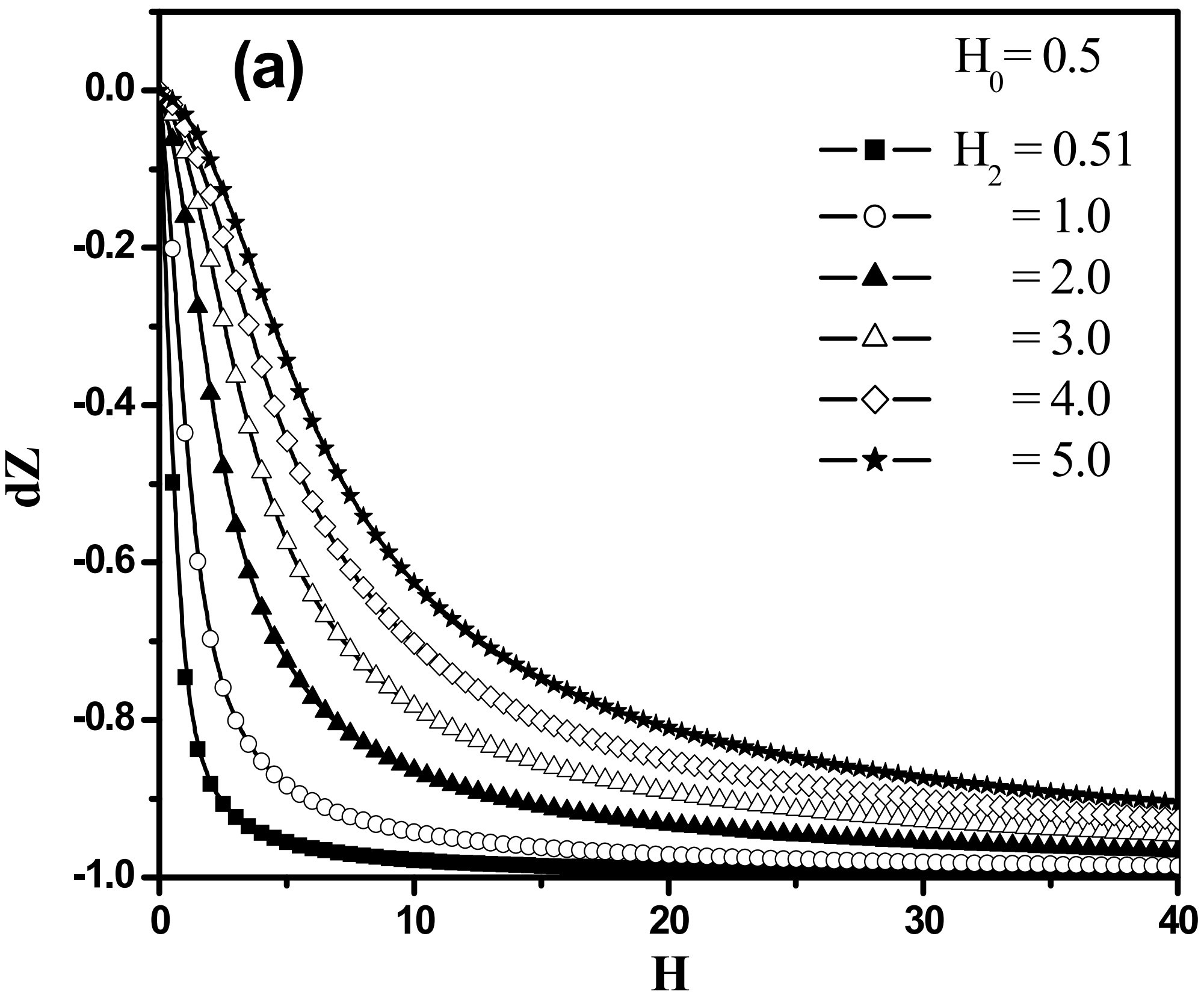

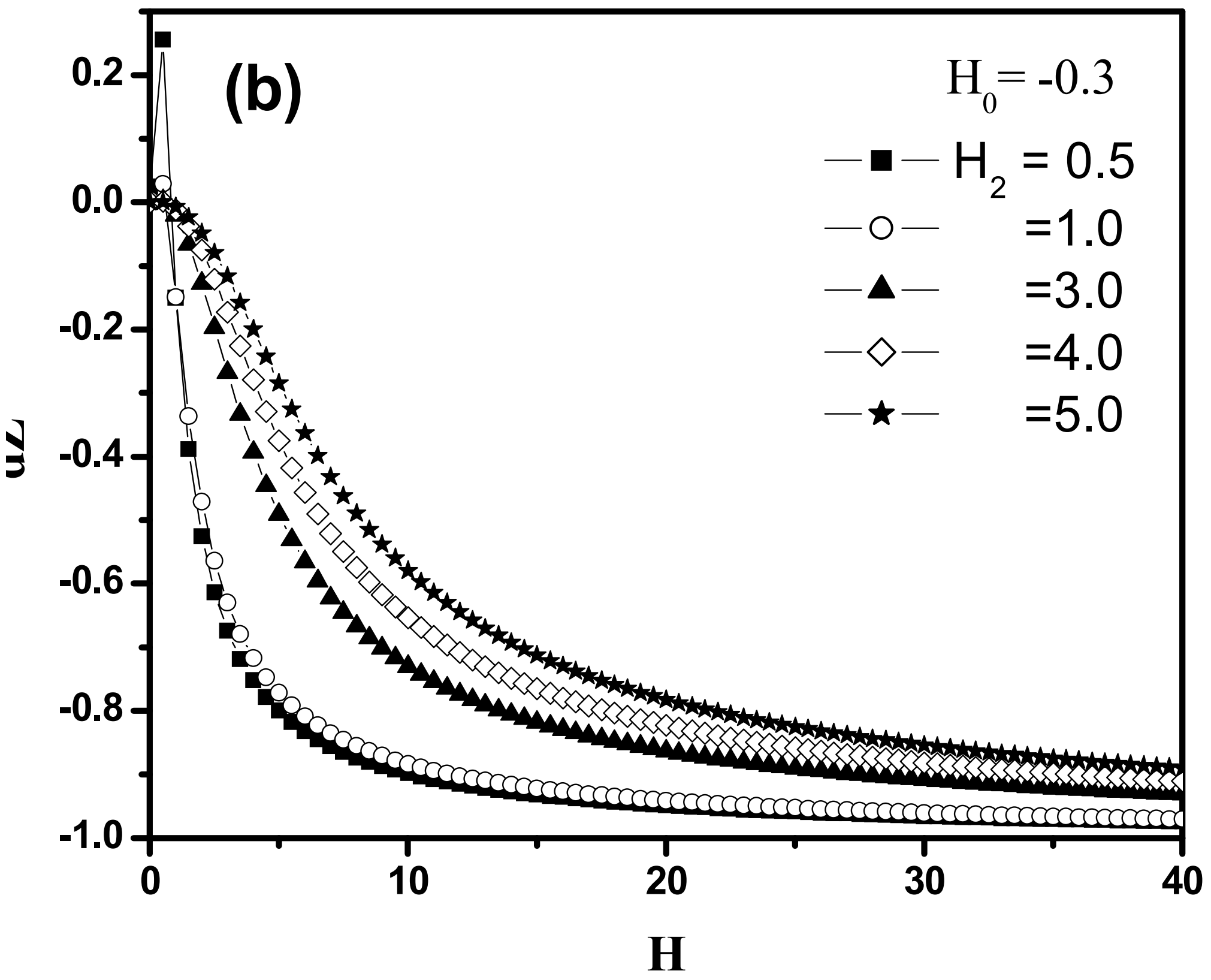

**FIG. 7 Ghatak et al.**

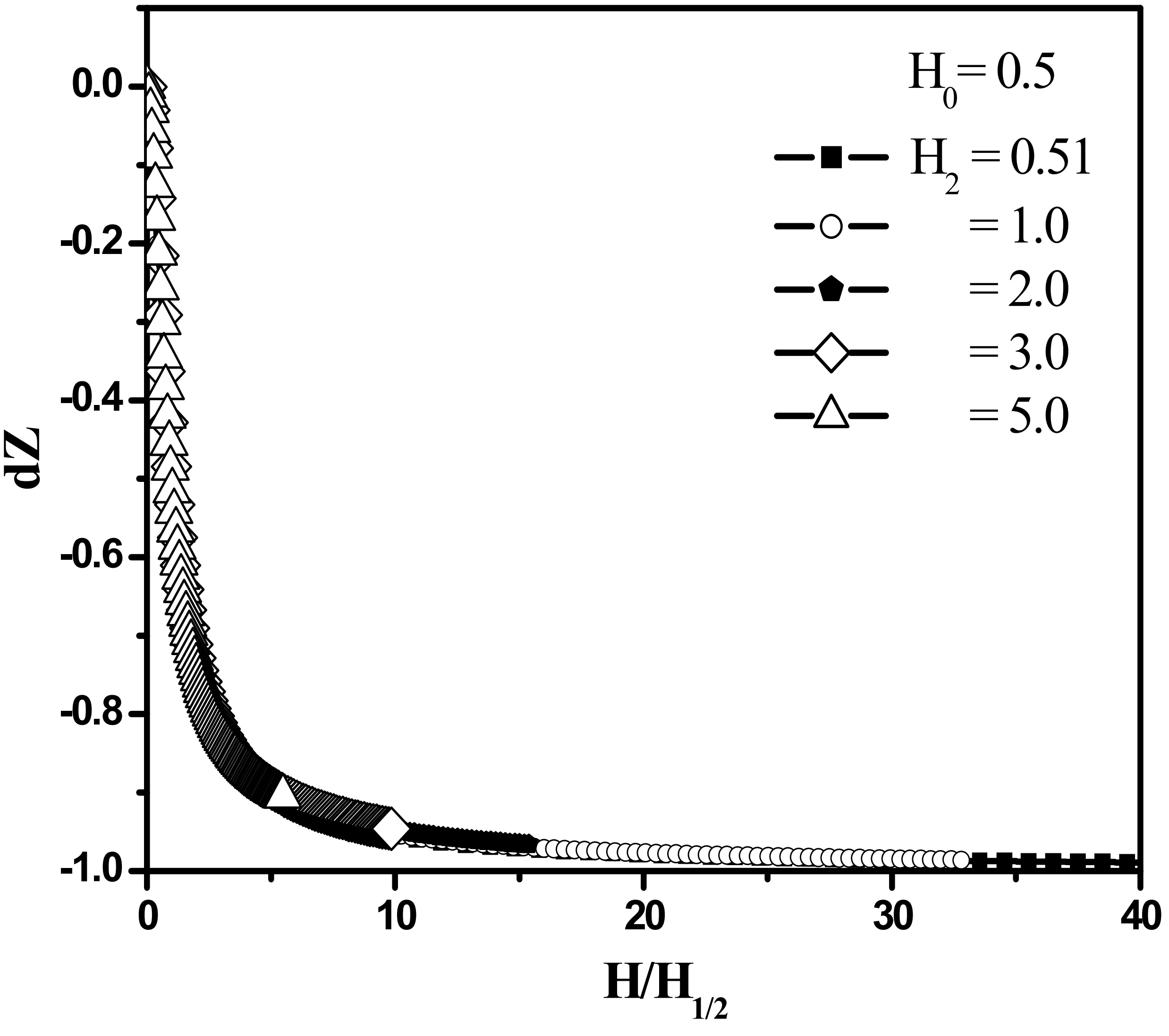

**FIG. 8 Ghatak et al.**

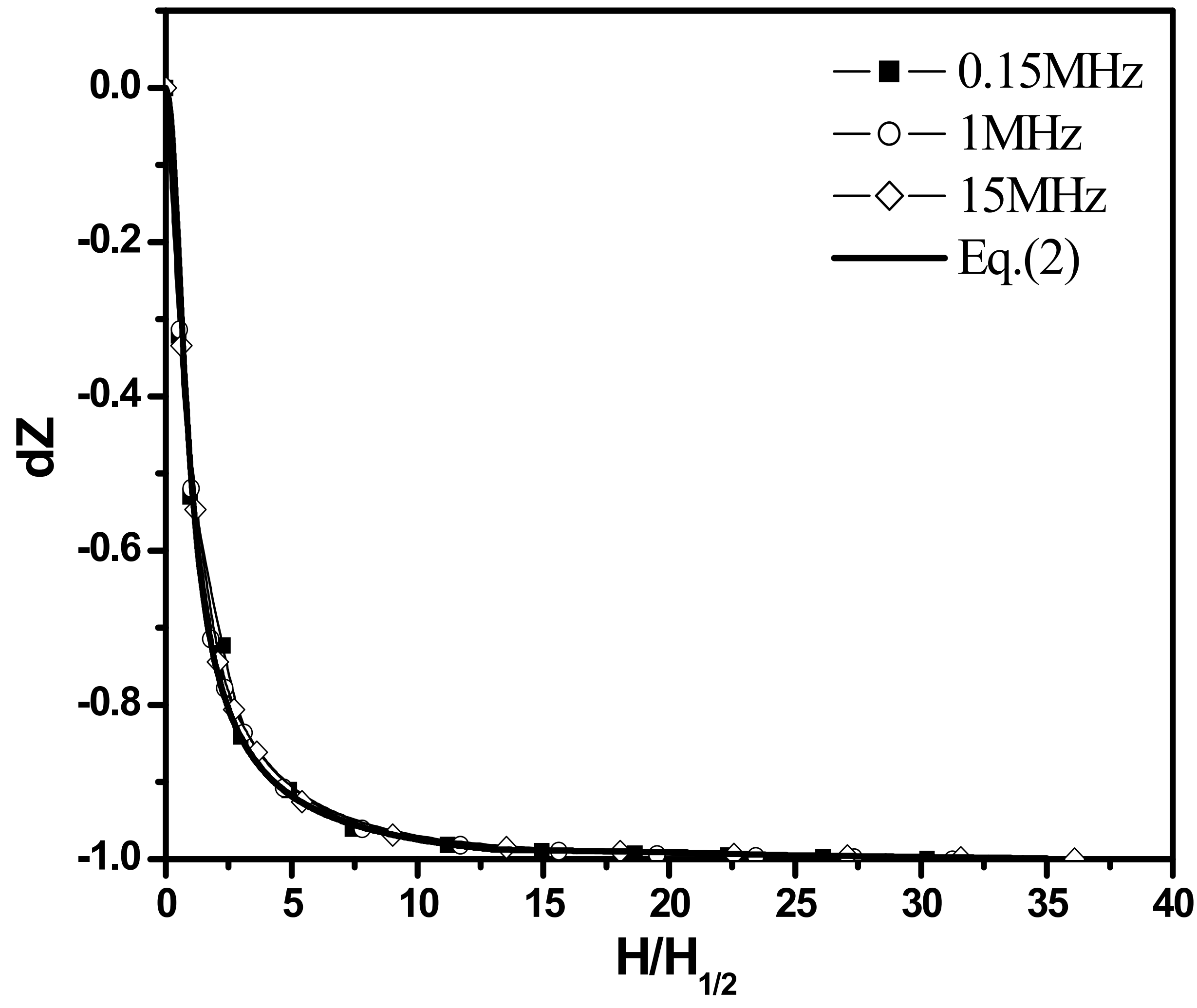

**FIG. 9 Ghatak et al.**